# High yield production of defect less carbon nanotubes in an arc process

*Soumen Karmakar[1], Naveen V. Kulkarni[1], V.G. Sathe[2], A.K. Shrivastava[3], M.D. Shinde[4], S.V. Bhoraskar[1,\*], and A.K. Das[5]*

[1]Department of Physics, University of Pune, Pune 411007, India

[2]UGC-DAE Consortium for Scientific Research, University Campus, Khandwa Road, Indore 452017, India

[3]Raja Ramanna Centre for Advanced Technology, Indore 452 013, India

[4]Center for Materials for Electronics Technology, Pune 411008, India

[5]Laser and Plasma Technology Division, Bhabha Atomic Research Centre, Mumbai 400085, India

Email: svb@physics.unipune.ernet.in

An efficient modified arc plasma method, where a focusing electric field is superimposed on the arc electric field, is optimized for the bulk generation of highly pure multi-walled carbon nanotubes (MWNTs). Raman spectroscopy and thermogravimetric measurements have been used to optimize the process. It was found that, at the optimized focusing field configuration, this process can utilize about 85% of the consumed anode material as compared to about 35% in the conventional arc plasma method. The sample prepared at the optimized conditions exhibited negligible D band intensity along with a reduced line width (14cm$^{-1}$) of the G band in the Raman spectrum. The oxidation temperature of this sample was found to be as high as 851°C. We conclude that an arc reactor can, thus, be modified suitably so as to achieve maximum yield and purity of the as synthesized MWNTs.

Carbon nanotubes are of great scientific interest due to their special microstructure and excellent physical, chemical and mechanical properties[1]. Right from its discovery[2], this novel material has been earmarked due to its great potential in a variety of scientific as well as industrial applications[3-8]. Arc plasma method is one of the earliest and very powerful techniques for generation of bulk quantity of carbon nanotubes[9]. However, arc-generated CNTs have inherited process-induced defects and it is necessary to follow a lengthy purification process[10] before using them in any kind of application. Arc process is also associated with a significant waste of the feedstock material in the form of amorphous carbon (a-C) and fullerenes[10, 11] during the process of synthesis. It is therefore of vital importance to explore ways to synthesize pristine CNTs with the maximum utilization of feedstock material.

We have reported earlier the effect of a focusing electric field on the formation of MWNTs in a graphite arc plasma reactor[12]. A more systematic investigation has led to optimize the electric field configuration to synthesize the best possible CNTs within the chosen geometry. The characteristic features of the as synthesized CNTs were subjected to detailed analyses using Raman spectroscopy and theromogravimetric measurements. High resolution transmission electron microscopy was useful to understand the multi-walled nature of the CNTs thus produced, at the optimized focusing conditions. The electrostatic focusing voltage ($\Phi$) was varied from 0V to 1200V, wherein a dramatic optimum behavior was observed near $\Phi$=400V.

Figure 1a shows the schematic of the modified reactor whereas Figure 1b depicts the dimensions and orientations of the electrodes' assembly used for the synthesis of MWNTs. The specific dimensions and orientations of the electrodes were designed in order to ensure zero leakage current through the focusing anode while maintaining a steady arc throughout the experiments. The details of the experimental procedures, sample preparations and the characterization equipments have been the same as mentioned elsewhere[12]. In addition, high resolution transmission electron micrographs were recorded using a Philips CM200 microscope.

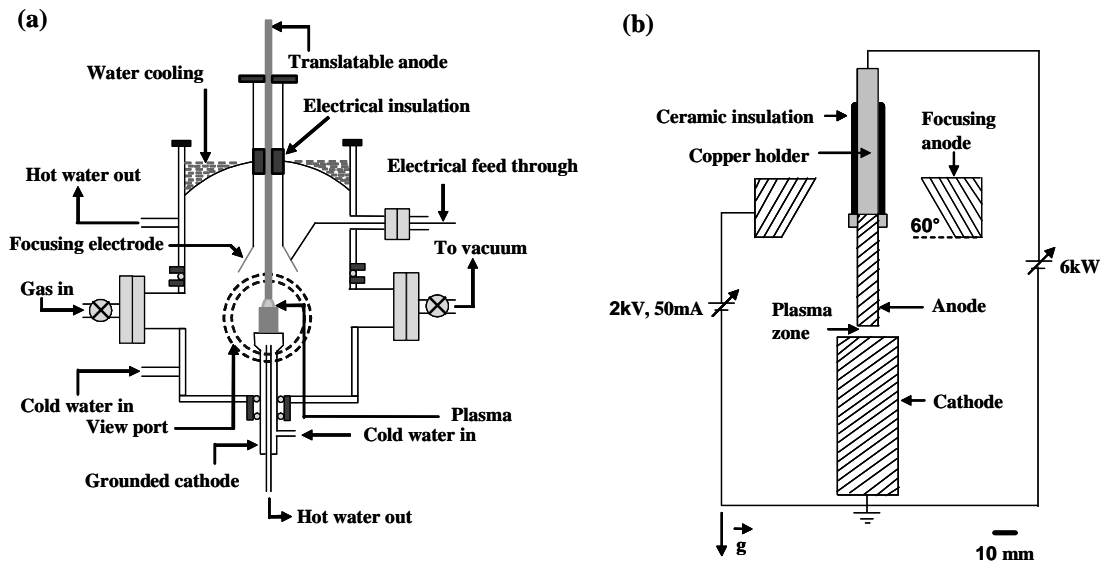

**Figure 1.** (a) Schematic of the modified electrode assembly, (b) dimensions and orientations of the electrodes' assembly.

The entire experiment was run at a steady current of 150A and voltage of 26V in an inert atmosphere consisting of He-H$_2$ (90% He, 10% H$_2$ by volume) mixture maintained at $6.66\times10^5$ Pa. The focusing electrode was a truncated cone-shaped graphite anode electrically insulated from the main electrodes. This was biased through a variable voltage regulated power supply (0-2kV rating). All applied voltages are referred with respect to the grounded cathode. Φ was varied from 0V to 1200V at a step of 200V; ensuring zero leakage current through the focusing electrode. The operational-time for each experiment was restricted to about 6 minutes.

The effect of symmetrically distributed focusing electric field on the formation of as synthesized MWNTs has been investigated, both by the microscopic and macroscopic characterizations of each sample, so as to correlate the results with Φ.

Mass-measurement has revealed results consistent with those reported in our earlier report[12]. The efficiency of cathode deposit formation ($\eta_{CD}$) is an important parameter, which is defined as $\eta_{CD}=$ [(mass of the cathode-deposit ÷ mass of the consumed anode material) ×100%]. Figure 2 shows the variation of $\eta_{CD}$ with $\Phi$. It is seen from Figure 2 that $\eta_{CD}$ initially increases with $\Phi$ rapidly from 0V to 400V reaching a saturation at $\Phi$=400V. From Figure 2 it is seen that most of the consumed anode material is deposited on the cathode-surface when the value of $\Phi$ is 400V or more. It is also noticed that $\Phi$=400V is enough to deposit most of the consumed anode material on the plasma facing surface of the cathode ($\eta_{CD}$≈85%). The increment in $\eta_{CD}$ for $\Phi \geq 400V$, as compared to the normal arc reactors is by ~170%. This is due to the reduction in the radial scatter and loss of the consumed anode material and has been discussed in detail elsewhere[12]. Figure 3 shows the Raman spectrum of the feedstock material used for the synthesis of MWNTs in the range $1200 cm^{-1}$ – $1700 cm^{-1}$. Figure 3a shows the first order Raman excitations of the feedstock glassy graphite like material and the samples prepared under different focusing bias configurations. He-Ne laser was used in our analyses as this laser has been reported to probe the presence of a-C and defective carbonaceous structures in an enhanced manner[16]. Here, for the ease of comparison, the amplitudes of all the G bands are normalized to unity after subtracting the backgrounds. The base lines of individual curves have been displaced arbitrarily. The areas under the peaks are referred to as the corresponding intensities. Common features in all these Raman spectra are the presence of two peaks. One, around $1330 cm^{-1}$, is correlated to the so called disorder induced D band and that around 1580cm$^{-1}$ to the so called G band[13]. G feature arises due to in-plane vibrations of the hexagonal graphene like structure[14], whereas the D band feature is best ascribed to the presence of defective sp$^2$ sites having de-localized π-electron clouds present within the samples[15]. From Figure 3a it is also seen that the D band for our feedstock material has a relatively higher amplitude and intensity than the G band. The peak position of G band is up shifted to a value of 1600cm$^{-1}$. The full width at half maximum ($W_G$) of this band is about 78 cm$^{-1}$. However, the spectra for the entire arc generated samples are quite different as the intensities of G bands are larger than those of D bands, the peak positions of G bands have down shifted to a value of around 1580 cm$^{-1}$ and the widths are around ≤26 cm$^{-1}$. A close examination of G band reveals its 'doublet' feature. A typical G band feature of the samples is therefore presented in Figure 4b as a magnified representative of all the samples with the characteristic two-Lorenzian fits. It is seen that the G band consists of two Lorenzians with their centers at 1580 cm$^{-1}$ and 1620 cm$^{-1}$ respectively. The peak at 1620 cm$^{-1}$ is a characteristic of defective graphite like structures and is smaller in better quality of MWNT containing samples[13]. While considering the G peak intensities, the contributions from the corresponding peaks arising at 1580 cm$^{-1}$ have only been considered.

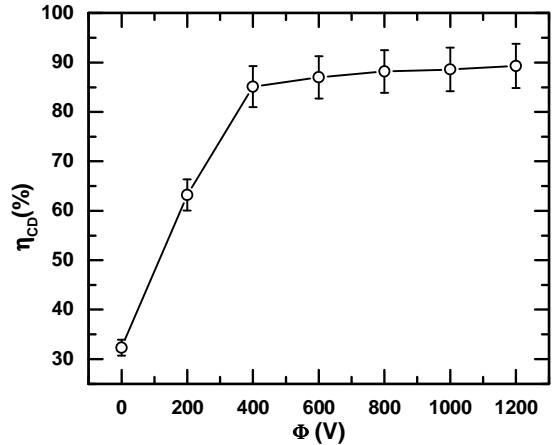

**Figure 2.** Variation of conversion efficiency of consumed anode material into cathode deposit.

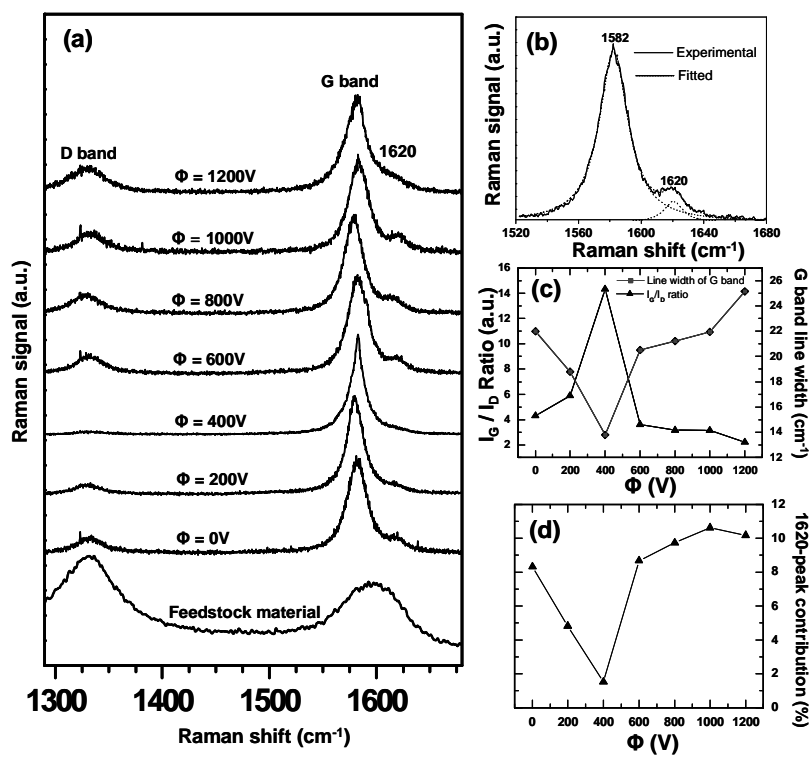

**Figure 3.** (a) Comparative Raman spectra of the samples synthesized under different Φ values at 632.81 nm laser excitations, (b) de-convolution of a typical G band into two Lorenzians, (c) variation of $I_G/I_D$ ratio and line width of G band with Φ, (d) percentage contribution of 1620- peak in the G band as a function of Φ.

The very first observation of these spectra indicates the remarkable differences in the relative intensities of D with respect to G bands as $\Phi$ is varied. The D band intensity has almost diminished at $\Phi$=400V and has increased when $\Phi$ is deviated from the value of 400V both for the lower and higher voltages. The relative intensity of the G band ($I_G$) to that of the D band ($I_D$) i.e. $I_G/I_D$ ratio is a good figure of merit in analyzing the fractional relative content of the ordered graphene structured species within a sample[17, 18]. Figure 4c shows the variation of this ratio with respect to $\Phi$ as derived from Figure 4a and b.

The response of $\Phi$ on the properties of the as synthesized MWNT containing samples are almost opposite with an extremum at $\Phi$=400V. Thus, we define two ranges of $\Phi$: $R_C \rightarrow$ 0V$\leq\Phi\leq$400V and $R_D \rightarrow$ 400V$\leq\Phi\leq$1200V respectively. It is seen that on increasing $\Phi$ within $R_C$, from 0V to 400V, $I_G/I_D$ ratio increases from a value of 4 to a value of 14, wherein the ratio drops down to a value of ~3 in $R_D$. An increase in the $I_G/I_D$ ratio indicates the decrease in the defect density and a-C content within a sample[19]. The behavior within $R_C$ is possibly due to the fact that up to $\Phi$=400V, application of $\Phi$ not only confines the carbon ions in the arc region, but also provides the requisite acceleration towards the cathode surface leading to the formation of pristine MWNTs. However, an opposite trend of variation of this ratio is seen in the region $R_D$. This indicates the introduction of defects within the samples with increasing $\Phi$. However, in spite of collecting and analyzing the total cathode deposits, the increase in the ratio at $\Phi$=400V further justifies the favorable conditions.

The variation of the G band line-width ($W_G$) with $\Phi$ is also noteworthy (Figure 4c). It has been reported[21] that the value of $W_G$ decreases on purifying a carbonaceous sample containing CNTs. It is seen that on applying $\Phi$, the value of $W_G$ falls down to a value of 14 cm$^{-1}$ at $\Phi$=400V from a value of 22cm$^{-1}$ corresponding to Φ=0V. It is noted from Figure 4c that on increasing Φ beyond 400V, $W_G$ steadily increases and reaches a value of 26 cm$^{-1}$ at Φ=1200V.

Figure 4d shows the variation of percentage contribution of the 1620 peak intensity in the G band as a function of Φ. From Figure 4d it is seen that in all the samples, excluding those at Φ=200V and 400V, the contribution of this peak is more than 8%. However, at Φ=400V this contribution is only ~1%.

Taken together the $I_G/I_D$ ratio, $W_G$ and the percentage contribution of 1620 peak intensity, it is inferred that at Φ=400V the CNTs are well graphitized and they pristine in nature. The corresponding sample does not contain much of a-C. Hence, Φ=400V corresponds to the optimized value of Φ where maximum purity of the as synthesized MWNTs can be achieved with about 85% utilization of the feedstock material. Our Raman results corresponding to Φ=400V are comparable to that reported[10] for the purified MWNTs.

In our previous work [12] we had observed radial breathing mode (RBM) at 279cm$^{-1}$. However, no RBM was observed for the present samples and it might account for the effects of the buffer gas, which was Ar-He mixture as against the He-H$_2$ mixture in the present.

Figure 4a shows a typical TEM micrograph of the CNTs synthesized at Φ=0V. A large amount of carbonaceous debris is well seen in the micrograph along with the CNTs.

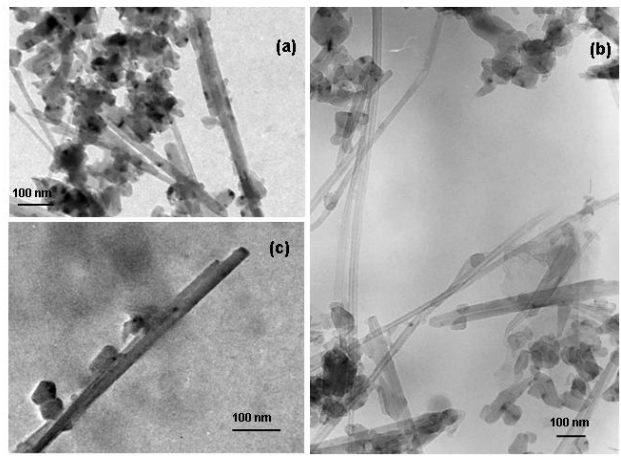

**Figure 4.** Transmission electron micrograph of dispersed cathode deposits synthesized at (a) Φ=0V, (b) Φ=400V and (c) Φ=1000V.

Figure 4b shows micrograph of the sample synthesized at Φ=400V. The CNTs synthesized at Φ=400V was found to have a narrow outer diameter distribution (10-25 nm). However, in Figure 4c, corresponding to Φ=1000V, heavy damages to the walls of the as synthesized MWNTs are clearly visible.

Figure 5a shows an isolated MWNT chosen from the sample prepared at Φ=400V. As this sample possesses maximum purity of pristine MWNTs as inferred from Raman spectroscopy, HRTEM micrographs were recorded only for this sample. Concentric circular planes of the MWNT are clearly seen in Figure 5b.

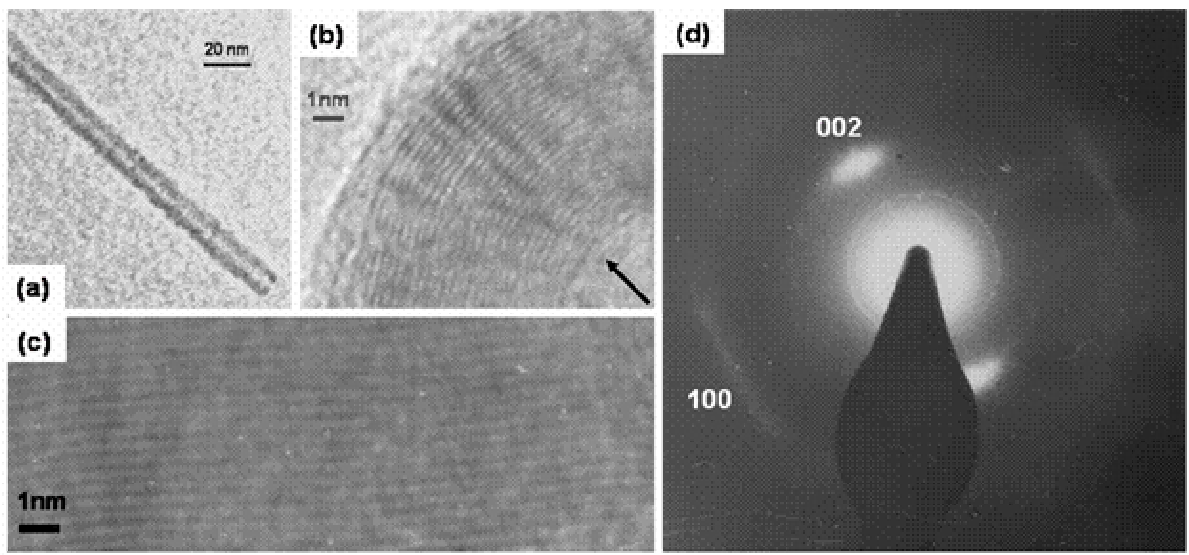

**Figure 5.** (a) TEM image of a typical tube synthesized at Φ=400V, (b) HRTEM image of a segment of cross-section of the tip of such an open-ended tube. Arrow is in the radially outward direction from the axis of the tube, (c) HRTEM micrograph of the side-wall of such a tube, (d) SAED pattern of such a tube.

In Figure 5c it is clearly seen that the walls are equally separated by an average distance of ~3.4Å and are parallel to each other. The crystallinity of these MWNTs was further studied by recording the Selected Area Electron Diffraction (SAED) pattern (Figure 6d). The pattern is in very good agreement with that reported in the literature[22]. The SAED pattern reveals the presence of (002) and (100) planes.

The oxidation behavior of CNTs is another important property and gives information about their crystalline purity and relative content in a carbonaceous sample. Figure 6a shows a typical thermogravimetric (TG) spectrum and its differential curve (DTG) of one of the samples. Similar TG-DTG spectra were recorded for all the samples (Figure 6a). The peak position of a DTG curve is referred to as the most probable oxidation temperature ($T_P$) for the corresponding sample. The temperature $T_P$ was derived from these spectra and is plotted as function of $\Phi$ in Figure 6b.

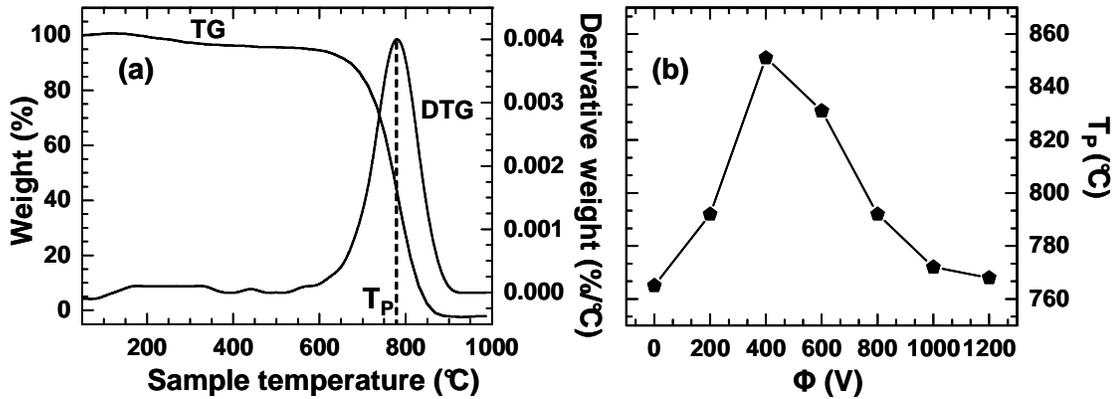

**Figure 6.** (a) Typical TG and DTG curves of one of our samples, (b) variation of most probable oxidation temperature ($T_P$) of the samples with $\Phi$.

It has been found that $T_P$ reflects the compositional aspects of any MWNTs containing sample and its purity level[12]. It has also been understood that the presence of carbon nanocrystalline particles (CNPs) and the defects, both tend to decrease $T_P$. Thus an increment in $T_P$ from a temperature of 762°C to 851°C on increasing $\Phi$ in $R_C$, shows that relative content of MWNTs and purity level in the corresponding samples have improved on application of $\Phi$. There is no doubt that the MWNT sample synthesized at $\Phi$=400V is certainly the best of the lot from both the 'relative content and purity level' points of view. A constant decrease in $T_P$ with increasing $\Phi$ in $R_D$ again confirms the fact that defects are indeed introduced and amorphous structures begin to dominate in the samples with increasing $\Phi$ in $R_D$.

Clubbing together both the analyses of Raman spectroscopy and thermal measurements it is hence reasonable to summarize the following:

(i) Within $R_C$, on increasing $\Phi$, formation of MWNTs is favored because of the generation of properly oriented axes of symmetry. Due to the appropriate strength and orientations of the focusing electric field, a favorable environment for the generation of MWNTs is created. This finding again supports the idea put forward in our earlier report[12].

(ii) Within $R_D$, there is no breaking of the axes of symmetry. However, the strength of the electric field crosses the threshold where it can introduce defects on the walls of the as synthesized MWNTs by *in situ* bombardment caused by the highly energetic charged plasma precursors. On increasing $\Phi$ within $R_D$, more and more defects are introduced in the tubes and this essentially leaves more sites for oxygen to react. As a result, a monotonous decrease in $T_P$ is found. Still, the ratio of numbers of MWNTs to that of CNPs remains higher than that at $\Phi=0V$. Otherwise, relatively higher values of $T_P$ in $R_D$ with respect to $\Phi=0V$ condition are hard to explain. There is need to carry out more careful investigation on the Raman scattering cross sections for the a-C, CNPs and defective MWNTs for estimating the relative content of MWNTs in a carbonaceous sample; since, $T_P$ shows an opposite trend, the values of $I_G/I_D$ ratio are smaller in $R_D$ than those in $R_C$.

In conclusion, we have optimized the synthesis method of multi-walled CNTs in which the arc electric field was superimposed with an additional electric field. This method, when optimized properly, yields highly pure CNTs minimizing the loss of consumed anode material in the form of a-C and/or fullerenes. One can, therefore, achieve bulk-productivity of highly pure MWNTs utilizing maximum of the consumed anode material (as high as ~85%) when the focusing voltage is tuned properly ($\Phi=400V$ for the present configuration) in this kind of modified reactor.

S. Karmakar would like to thank BARC-PU Collaborative Research Program for providing financial support. Acknowledgement is due to Dr. R.V. Nandedkar of RRCAT, Indore, India for the use of HRTEM.